# Direct visualisation of dislocation dynamics in grain boundary scars

Peter Lipowsky[1], Mark J. Bowick[2], Jan H. Meinke[2], David R. Nelson[3] and Andreas R. Bausch[1]

[1]Lehrstuhl für Biophysik E22, TU München, 85747 Garching, Germany

[2]Physics Department, Syracuse University, Syracuse, New York 13244-1130, USA

[3]Lyman Laboratory of Physics, Harvard University, Cambridge Massachusetts 02138, USA

**Mesoscale objects with unusual structural features may serve as the analogues of atoms in the design of larger-scale materials with novel optical, electronic or mechanical behaviour. In this paper we investigate the structural features and the equilibrium dynamics of micron-scale spherical crystals formed by polystyrene particles adsorbed on the surface of a spherical water droplet. The ground state of sufficiently large crystals possesses finite-length grain boundaries (scars). We determine the elastic response of the crystal by measuring single-particle diffusion and quantify the fluctuations of individual dislocations about their equilibrium positions within a scar determining the dislocation spring constants. We observe rapid dislocation glide with fluctuations over the barriers separating one local Peierls minimum from the next and rather weak binding of dislocations to their associated scars. The long-distance (renormalised) dislocation diffusion glide constant is extracted directly from the experimental data and is found to be moderately faster than single particle diffusion. We are also able to determine the parameters of the Peierls potential induced by the underlying crystalline lattice.**

There is considerable need for the rational design of new functional materials. One promising strategy is to build such materials from the bottom up by assembling mesoscale units which can then be linked into larger structures. The characteristics of such materials on the fundamental scale of the building blocks must be understood before they can be intelligently assembled into novel materials. In many cases the units in question possess surfaces with ordered arrays of particles; liposomes, colloidosomes, fullerenes and nanotubes provide prominent examples of such materials[1-3]. As in traditional 3D materials, the static structure and dynamic behaviour of defects, such as dislocations, is crucial in determining the response to mechanical, electrical and thermal stimuli[4]. In the ground state of planar 2D systems (flat space) all such defects are tightly bound. The role of thermally excited or mechanically induced defects has been thoroughly studied from both the theoretical and experimental side[5,6].

Much less is known, however, about defect dynamics since it is difficult to directly observe the time evolution of defects in most experimental systems. In the technologically relevant case of curved surfaces it is known that new defect structures arise even at zero temperature. Dislocations form in the ground state of sufficiently large curved crystals because they lower the total elastic energy[7]. In the specific case of spherical crystals these dislocations may be viewed as screening the elastic strain of the isolated disclination defects required by the topology of the sphere. They are present, above a critical particle number of order 300, in the form of novel freely terminating high-angle grain boundaries dubbed scars[8].

While the structural features of spherical crystals are very rich, we concentrate here on the dynamics of particles crystallizing on a spherical surface as well as the dynamics of the resultant dislocation arrays. We visualise and track the diffusion of both single particles and dislocations as they fluctuate about their equilibrium positions and determine the dislocation spring constant. The presence of dislocations in the ground state enables us to determine the long distance dislocation diffusion constant analyzing the observed Peierls potential.

Spherical crystals can be characterised by their dimensionless system size *R/a*, where *R* is the radius of the sphere and *a* is the lattice constant determined by the mean particle spacing. For *R/a* bigger than about five, the minimal energy configurations contain 12 extended arrays of disclination defects, each with net defect charge +1, rather than 12 isolated charge +1 disclinations. Provided *R/a* is not too large, the structure of each defect array is typically a linear chain of alternating 5 and 7-fold coordinated particles with one excess 5. A tightly bound 5-7 pair is itself a point-like topological defect in two dimensions known as a dislocation. Dislocation lines in three dimensions are important in determining the properties of bulk materials and play a crucial role in plastic deformations[4]. In flat two-dimensional systems their formation at finite temperature drives the melting of crystals to hexatics[5,6]. A line of dislocations is a grain boundary, characterised by the change in the orientation of crystallographic axes as one crosses the grain boundary. Since dislocations have vanishing total disclination charge there can be an arbitrary number of them in any spherical lattice configuration without violating the topological constraint on the total disclination charge discussed above. The presence of excess dislocations beyond the disclinations demanded by the topology of the sphere may be viewed as an elastic analogue of Debye screening. Dislocations screen the elastic strain energy of isolated disclination defects, which otherwise grows quadratically with system size.

The grain boundaries found in spherical crystals are unlike any found in flat space as they terminate freely inside the crystal at both ends. In flat space, grain boundary free ends are highly suppressed energetically because of the resultant clash of crystallographic orientations, whereas the constant positive Gaussian curvature of the sphere allows a finite number of excess dislocations to lower the elastic strain energy of an isolated 5 disclination for sufficiently large radius *R/a*. The diffusive behaviour of mono or divacancies in flat space has been observed by explicitly creating vacancies in a 2D colloidal crystal using optical tweezers. These fundamentally different defect structures are dislocation multipoles with zero net Burgers vector and were observed to diffuse in an attractive potential modulated by a Peierls potential[9,10]. These authors did not, however, determine the renormalised diffusion constant.

Much less is known about the dynamics of particle ordering on curved surfaces or the dynamics of dislocations. The experimental model system we study is formed by solid spherical colloidal beads adsorbed at a liquid-liquid interface. These so-called Pickering emulsions are of technological importance as well as powerful model systems for understanding ordering and dynamics in two dimensions[11,12]. The specific system we explore consists of water droplets dispersed in toluene; the colloidal particles are divinylbenzene cross-linked polystyrene microspheres with a diameter of about 1 μm. The microspheres have the appropriate interfacial surface tensions to adsorb at the oil-water interface. They crystallise once their areal density is sufficiently high. The carboxylate-modified surface of the microspheres has a negative surface charge at neutral to high pH, resulting in a screened Coulomb repulsion which prevents aggregation. We typically observe water droplets with a diameter of 20 to 100 μm. Microsphere-coated droplet images were obtained with optical bright field microscopy using an inverted microscope (Axiovert 200, Zeiss). The images were captured by a CCD-camera (Orca ER, Hamamatsu) with frame rates of 20 to 100 Hz and written directly to hard disk with digital image processing software[13]. We determined the particle positions using a standard tracking routine which fits a 2D Gaussian to the intensity profile of each particle. The spherical curvature of the droplets limited the imaged surface area to between 10 and 25 % of the crystal. To correct for drift motion, the centre of mass of all tracked particles was calculated for each frame and subtracted from the single particle coordinates before further analysis of the data. For each frame the lattice was triangulated automatically with a Delaunay algorithm appropriate to the sphere[14]. Disclinations were highlighted and their positions were tracked.

We found that small droplets have only isolated 5-fold (+1) disclinations. Droplets with *R/a* above 5, however, exhibit scars. Although scars are expected to follow geodesics on the sphere at zero temperature, thermal fluctuations at room temperature bend them significantly. A single scar with a kink is shown in **Fig. 1**. The fixed angular length of scars implies that the number of excess dislocations per scar should grow linearly with *R/a*. The predicted slope, of order 0.41, is independent of the microscopic potential and compares favorably with experimental observations[8].

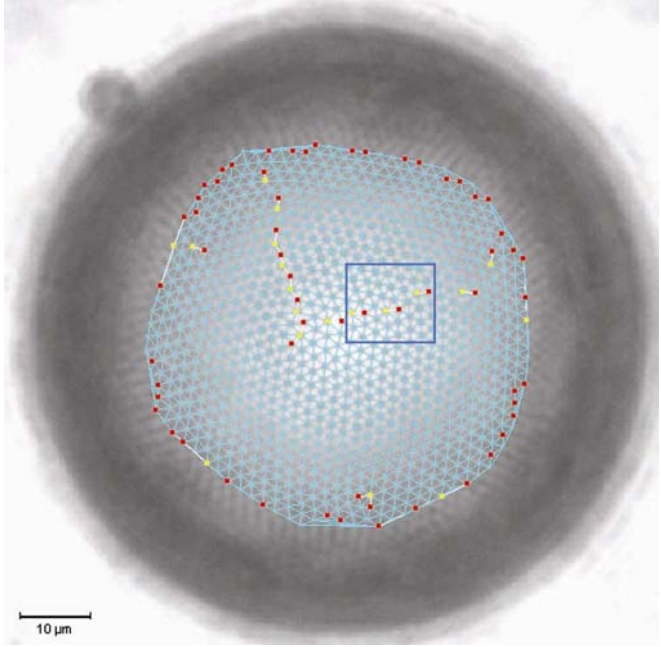

**Fig. 1**: A light microscope image of an 85 μm diameter water droplet coated with 1 μm diameter polystyrene microspheres. The mean spacing is 1.9 μm ($R/a \approx 22$). About 11 % of the crystalline lattice is triangulated (835 particles). 5-fold coordinated particles (+1 disclinations) are coloured red and 7-fold coordinated particles (-1 disclinations) are coloured yellow. A single bent grain boundary (scar) is clearly evident. The blue box encloses the three dislocations whose motion is displayed in Figs. 4 and 5. The left-most of these three dislocations is only one dislocation away from the centre of the scar.

What is the elastic response of our spherical colloidal crystals? Although macroscopic methods, such as AFM indentors, can be used to determine elastic constants, their force resolution is not good enough to study very soft materials[15]. Single-particle diffusion measurements, on the other hand, are ideally suited for soft materials such as colloidal crystals[5]. Since dislocations move by particle rearrangement, a quantitative understanding of single particle dynamics is also an essential prerequisite to understanding the dynamics of dislocations.

A three parameter model adequately describes the mean squared displacement of single particles in a colloidal crystal[16]. On short time scales the colloids diffuse freely, as characterised by the diffusion constant *D*. The restoring potential provided by the (harmonic) interaction with neighbors soon limits this linear growth, which eventually saturates. Additional diffusive motion is also expected due to plastic deformation of the lattice caused by the diffu-sion of thermally excited free defects. These three contributions to particle motion are evident in the mean-square displacement:

$$\langle \Delta x^2 \rangle = \tfrac{1}{2} \cdot \frac{\dfrac{k}{k_B T} + \dfrac{1}{4 \cdot D \cdot t}}{\left(\dfrac{k}{2 \cdot k_B T} + \dfrac{1}{4 \cdot D \cdot t}\right)^2} + 2 \cdot D_d \cdot t, \qquad (1)$$

where $\Delta x$ is the displacement of the particle from its equilibrium position along one direction $x$ in the crystal, $k$ is the effective spring stiffness of the harmonic lattice potential, and $D_d$ is the diffusion constant associated with defect-mediated diffusion.

In **Fig. 2** we show the measured mean squared displacement of single colloidal particles. To improve statistics the mean squared displacement was averaged over particles in patches varying in size from 90 to 200 particles. A standard regression fit yields $D \approx 0.02\ a^2/s$, $k \approx 145\ k_B T/a^2$ and $0 < D_d < 2 \cdot 10^{-4}\ a^2/s$. Since the typical mean particle spacing is $a = 2$ μm, this translates to a value $D \approx 0.08$ μm²/s. This is lower than for diffusion in bulk water ($D \approx 0.4$ μm²/s), probably due to interfacial hydrodynamic effects[17].

For an isotropic 2D crystal with nearest-neighbour interactions the Young modulus Y is given by $Y = \dfrac{2}{\sqrt{3}} \cdot k \approx 167\ k_B T/a^2 \approx 1.7 \times 10^{-7}$ N/m. The compressibility modulus $K = (3/4) \cdot Y \approx 125\ k_B T/a^2$; this is of the same order of magnitude as those measured in three-dimensional colloidal crystals[16].

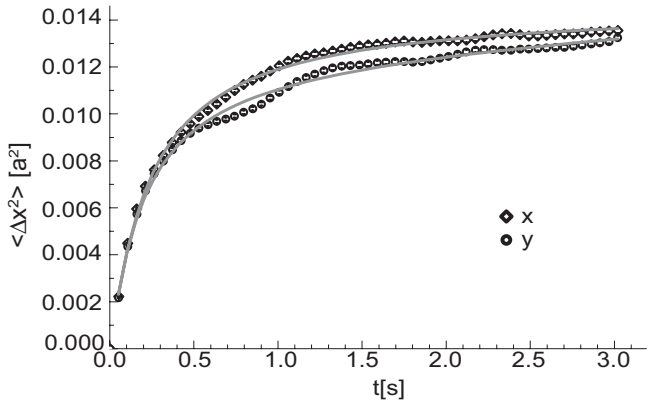

**Fig. 2**: Mean squared displacement of single colloidal particles. The averaging over many time steps as well as over 159 particles yields statistical error bars smaller than the symbol size. Fits to Eq. (1) give $k_x = 136\ k_B T/a^2$, $D_x = 0.02\ a^2/s$ and $D_{dx} \sim 0$ along the x-axis; $k_y = 157\ k_B T/a^2$, $D_y = 0.02\ a^2/s$ and $D_{dy} \sim 2\ 10^{-4}\ a^2/s$ along the y-axis.

The motion of dislocations in materials involves oscillations in the local minima of the periodic (Peierls) potential due to the underlying crystalline lattice as well as thermal fluctuations over the barriers separating local minima. Dislocation motion can be separated into glide and climb. Glide is motion parallel to the dislocation's Burgers vector or perpendicular to the axis of the dislocation, and requires only a local rearrangement of the lattice. Climb is motion perpendicular to the Burgers vector or parallel to the axis of the dislocation and requires the presence of vacancies or interstitials. The diffusion constants for glide are therefore expected to be much higher than for climb.

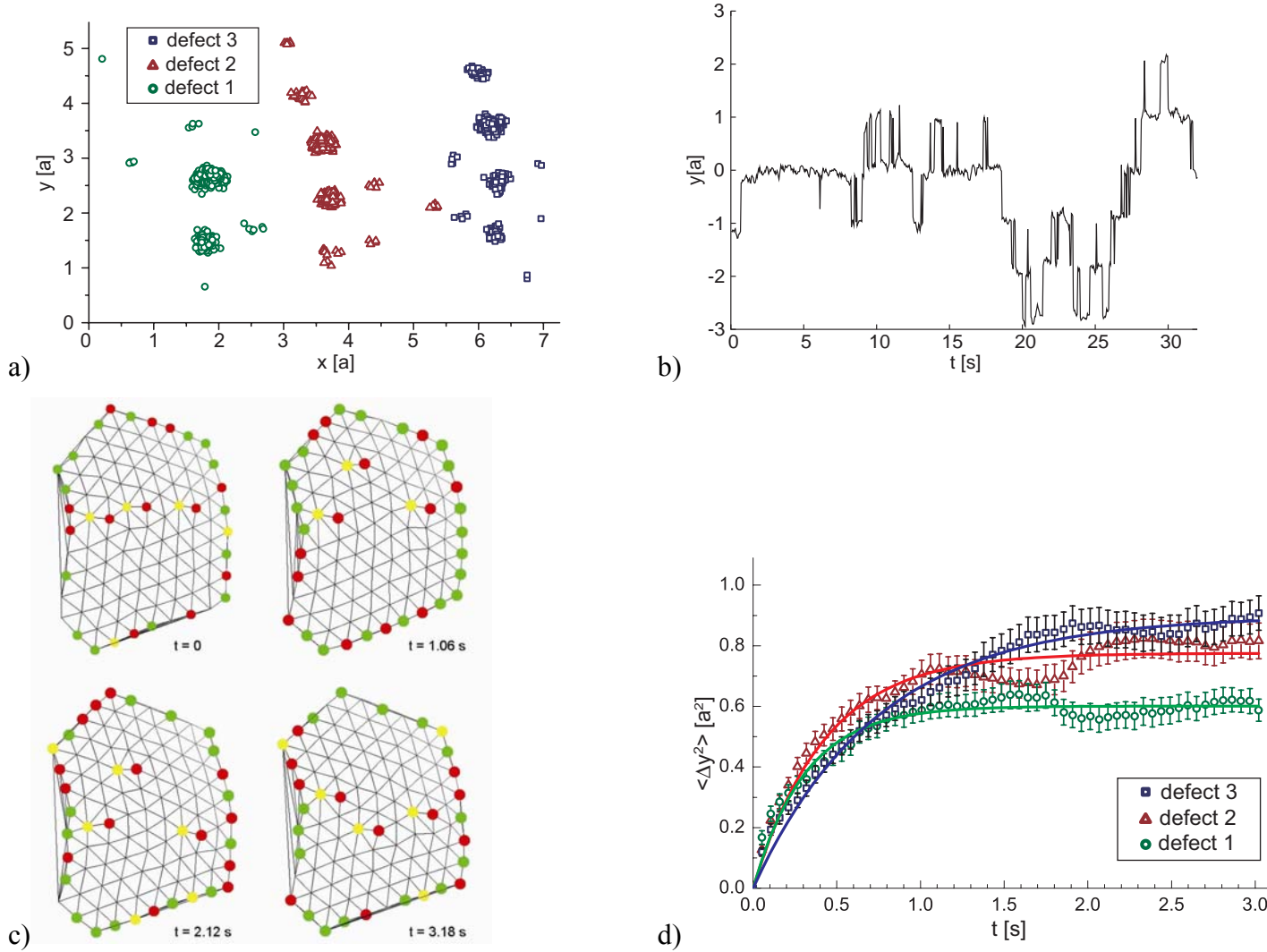

Fig. 3: Visualisation and evaluation of the motion of dislocations bound to a grain boundary scar. (a) Plot of each dislocation's motion in the x-y plane. (b) Glide motion of a single dislocation. (c) Motion of three dislocations gliding about their equilibrium positions within a grain boundary scar. The highlighted lattice points on the margin of the patch are boundary artefacts. (d) Comparison of the mean squared glide displacements for the three distinct dislocations boxed in Fig. 1. The central dislocation is more strongly bound than those near the end of the scar, as confirmed by fits to Eq. (2):
$k_1 = 1.7\ k_BT/a^2$, $k_2 = 1.3\ k_BT/a^2$ and $k_3 = 1.1\ k_BT/a^2$. The diffusion constants are similar for all three dislocations: $D \approx 1\ a^2/s$. Error bars show the standard deviation of the calculated mean-square displacements.

A plot of the motion in the x-y plane for three distinct dislocations in a single grain boundary scar is shown in **Fig. 3a**. Glide in the y-direction is clearly visible while climb is minimal. A plot of the glide motion along the y-axis, as a function of time, for a single dislo-cation fluctuating about its equilibrium position in a scar is shown in **Fig. 3b**. The plateaus correspond to oscillations about a fixed local minimum of the Peierls potential and the discrete jumps by a lattice spacing correspond to thermal fluctuations over a Peierls barrier to a new local minimum. Finally we display in **Fig. 3c** four snapshots of the instantaneous locations of three dislocations within a single scar at time intervals of 1.06 seconds. A movie version of **Fig. 3c** is available in the supplementary material.

The mean squared displacements of all three dislocations are plotted in **Fig. 3d**. At short times ($t < 0.5$ s), the dislocations exhibit Brownian diffusion within a local well while at longer times the mean squared displacement saturates due to the potential binding each dislocations to the scar as a whole. For small deflections $y$ of the dislocation in the glide direction the binding potential can be shown to be harmonic with a spring constant $k_D$. Modeling glide by a Langevin equation in the glide coordinate $y$, we obtain the mean squared displacement of a dislocation in a defect scar:

$$\langle \Delta y^2 \rangle = \frac{k_B T}{k_D}\left(1 - e^{-2\mu \cdot k_D \cdot t}\right), \quad (2)$$

where $\mu$ is the mobility of the dislocations. We note here that the long-time asymptote of Eq. (2) differs from that in Eq. (1) because we are not averaging over random initial locations of a dislocation. A simple fit to the data gives $D^0_{glide} = \mu\, k_B T \approx 1\ a^2/\text{s}$. This is roughly two orders of magnitude larger than the diffusion coefficient of single particles. The rapid Brownian diffusion of defects (vacancies and di-vacancies) has also been observed in flat space colloidal experiments[9]. The dislocation diffusion constants we obtain are a factor 3 smaller than those obtained for the diffusion of vacancies in flat space in Ref. 16. A vacancy, however, does not have a very well-defined identity. It is realised as a complex array of dislocations and diffuses quite differently from individual dislocations with a net Burgers vector.

The true displacement of a dislocation is determined not by its free diffusion constant, $D^0_{glide}$, which describes only small diffusive fluctuations, but by a much more physical renormalised diffusion constant $D^{ren}$ that includes the effects of large displacements by fluctuations over barriers. The renormalised diffusion constant is that which determines the time needed for the formation of equilibrium structures such as scars since this involves large scale motions of dislocations as they rearrange to screen the elastic strain energy of isolated disclinations. We are able to extract this more physical diffusion constant directly from our experimental data. It has been established in the theory literature that the two diffusion constants are related as follows:

$$D^{ren} = D^0_{glide} / I^+ I^-,$$

where $I^{\pm} = \int_0^a \frac{dy}{a} e^{\pm U_P(y)/k_B T}$ and $U_P(y)$ is the Peierls potential[18,19]. By constructing the probability distribution of a diffusing dislocation as a function of glide coordinate it is possible to extract $D^{ren}$ unambiguously. Since the glide histogram gives us $N(y) \propto e^{-(U_{tot}(y)/k_B T)}$, with $U_{tot}(y) = \frac{1}{2} k y^2 + U_P(y)$, we only need to subtract off the best fit Gaussian and numerically integrate to obtain $I^{\pm}$ (Fig. 4). We take for $U_P(y)$ a simple cosine potential $U_P(y) = -U_0 \cos(2\pi y / a)$, for which $I^+ = I^-$. We thus find $D^{ren} \approx D^0_{glide} / 8 \approx 0.1$ $a^2$/s. Free dislocations are thus able to migrate over a distance R in a time $t \approx R^2 / 4D^{ren} \approx 20$ min, providing ample equilibration time over the time scales of observation. Fitting the histogram yields $U_0 \approx 2.3$ $k_B T$ and $k \approx 2.5$ $k_B T/a^2$. The full harmonic plus Peierls potential extracted directly from the data is shown in the inset of **Fig. 4**. The symmetry of the potential underlines the fundamentally different nature of the dynamic behaviour of screening dislocations on curved surfaces compared to the dynamics of vacancies in flat space.

It is a general feature of the elastic Hamiltonian governing dislocations that they are bound to grain boundaries. This is well known in flat space[20] where the binding potential is harmonic for small displacements with an effective spring constant $Y\pi^3/432 \sim 0.07$ $Y \sim 14$ $k_B T/a^2$. On the sphere we have the additional feature that the grain boundary scars themselves have a net disclination charge. Near the end of a scar the disclination charge of the scar itself is screened completely by the Gaussian curvature of the sphere and by the dislocations within a scar. An estimate of the effective spring constant $k_D$ can be made using the defect potential derived in the continuum theory[7]. The strength of the potential decreases with distance from the centre of a scar with maximum values for the spring constants similar to those in flat space. From our dislocation diffusion data we find values for the dislocation spring constants $k_D$ between 0.3 and 1.7 $k_B T/a^2$. Thus dislocations are much more weakly bound than theory indicates. In qualitative agreement with predictions we observe the strongest spring constant for the innermost dislocation.

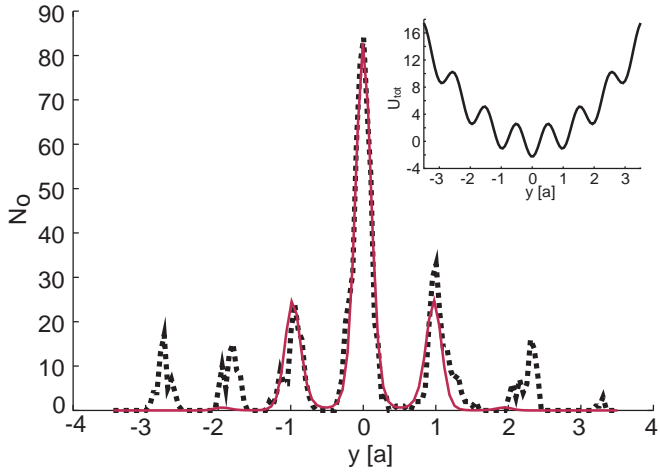

Fig. 4: Plot of the probability distribution of a diffusing defect as a function of position (dotted line). The red line indicates the fit to a harmonic potential modulated by the Peierls potential, as shown in the inset.

The observed weak binding may be due to effective softening of the spring constant by thermal fluctuations which is expected to be significant in our room temperature experiment. It may also be necessary to take account of the thermal bending of scars as a whole – in general we note that scars undergo significant shape fluctuations during the time scale of dislocation glide and this surely leads to an underestimate of the spring constant. Further studies should investigate potential dynamic collective phenomena in dislocation dynamics and their influence on material properties.

The direct optical observation of particle dynamics on curved crystalline surfaces allowed us to determine the material elastic response. We also quantified the parameters governing dislocation motion, including the renormalised diffusion constant governing the large distance motion of dislocations across the sphere, the harmonic potential binding dislocations to grain boundary scars and the periodic Peierls potential due to the underlying crystalline lattice. The dislocations assemble into scars; they are highly mobile perpendicular to the scars and bound more weakly than expected. The observed defect scars may be harnessed as functional sites in inorganic or biological systems but may also be regions of weakness and sources of plastic flow. The dynamic nature of dislocations in grain boundary scars will have to be accounted for in the design of novel nano and mesoscale materials.

**Acknowledgements**

This work was supported by the DFG (BA2029/5) and partly by the Fonds der chemischen Industrie. The work of MJB and JHM was supported by the National Science Foundation through Grant No. DMR-0219292 (ITR). The work of DRN was supported by the National Science Foundation through the Harvard Material Research Science and Engineering Laboratory via Grant No. DMR-0213805 and via Grant No. DMR-0231631. We thank Jörg Schilling for the help he provided with imaging processing and Angelo Cacciuto and James McCullough for the Java Applets containing the triangulation routines. We are grateful for valuable discussions with Alex Travesset and M. Nikolaides.

Correspondence and requests for materials should be addressed to A. R. B. (abausch@ph.tum.de) or M. J. B. (bowick@physics.syr.edu)